# Shor 演算法實作：在特定條件限制下分解 4096-Bit 整數

Abel C. H. Chen
Information & Communications Security Laboratory,
Chunghwa Telecom Laboratories
Taoyuan, Taiwan
ORCID: 0000-0003-3628-3033

**摘要—近幾年隨著量子晶片 Willow 等技術的發展，降低量子計算錯誤率，將有機會帶動量子計算進入實務應用的場域。因此，設計可以實務應用的量子演算法將會是重要的研究方向之一。在本研究中，主要聚焦在 Shor 演算法實作，改進模數計算效率，並且展示可以在特定條件限制下分解 4096-Bit 整數。在實驗結果中與最先進的(State-of-the-Art, SOTA)方法比較，證明可以提升數十倍效率，並且可以做到分解長度更長的整數。**

## I. 前言

在 2024 年 12 月，Google 研究團隊在《Nature》上發表論文“ Quantum error correction below the surface code threshold”，提出量子晶片 Willow，可以有效提供量子糾錯能力，降低錯誤率，從而帶動量子計算的發展[1]。因此，盤點現有的量子演算法，包含量子搜尋演算法 Grover 演算法[2]、量子質子因分解 Shor 演算法[3]等，嘗試把現有的量子演算法落實到真實應用場域將會是現階段的重要研究主軸之一。其中，Shor 演算法主要包含模數計算(Modular Computation)和逆量子傅立葉變換(Inverse Quantum Fourier Transform, IQFT)計算兩個部分，在質因數分解上可以透過量子計算特性來實現指數級加速，將可能威脅現行主流密碼學 RSA-2048 的安全性[4]。

有鑑於此，本研究在 Shor 演算法上進行改進，提出更有更高計算效率的模數計算量子電路，並且嘗試在特定條件限制下分解長度更長的整數。在符合特定條件限制下，本研究的主要貢獻條列如下。

1. 提出高效率的模數計算量子電路，與最先進的(State-of-the-Art, SOTA)方法[5]-[10]比較，證明可以提升二十倍的效率。
2. 本研究方法可以用更少的量子位元數來分解長度更長的整數，包含分解 4096-Bit 整數。
3. 本研究採用 IBM Qiskit 實作量子電路和量子演算法，通過實作來驗證本研究提出方法的可行性及其效率。

本研究主要分為五個小節。第 II 節介紹 Shor 演算法的基本流程，並且第 III 節介紹本研究提出的改進模數計算量子電路，並且搭配實例說明。第 IV 節說明實驗環境，並且與最先進的方法進行比較，提供比較討論。最後，第 V 節總結本研究貢獻，並且討論未來可行的研究方向。

## II. SHOR 演算法

本節將先說明 Shor 演算法的基本流程，再以近年文獻發表的例子來做討論。

### A. 演算法流程

圖 1 為 Shor 演算法的基本流程。假設要分解的整數是 $n$-bit 整數 $N$ (即$N \le 2^n$)，並且隨機選擇一個整數 $a$ 作為底數(在圖 1 中$a = 2$)。其中，$p$ 和 $q$ 各別為質數，$N = pq$，$a$ 和 $N$ 的最大公因數必須為 1。產生兩個註冊器(register)各包含 $k$ 個量子位元，總共 $2k$ 個量子位元。其中，$q_0$~$q_{k-1}$ 為第 1 個註冊器的量子位元，$q_k$ ~ $q_{2k-1}$ 為第 2 個註冊器的量子位元，並且在後續計算在第 1 個註冊器和第 2 個註冊器之間存在量子糾纏態。在計算上，主要包含三個部分：模數計算、逆量子傅立葉變換計算、取得週期 $r$ 和分解因數，可以用來分解 $n$-bit 整數(其中$n \le k$)，分述如下。

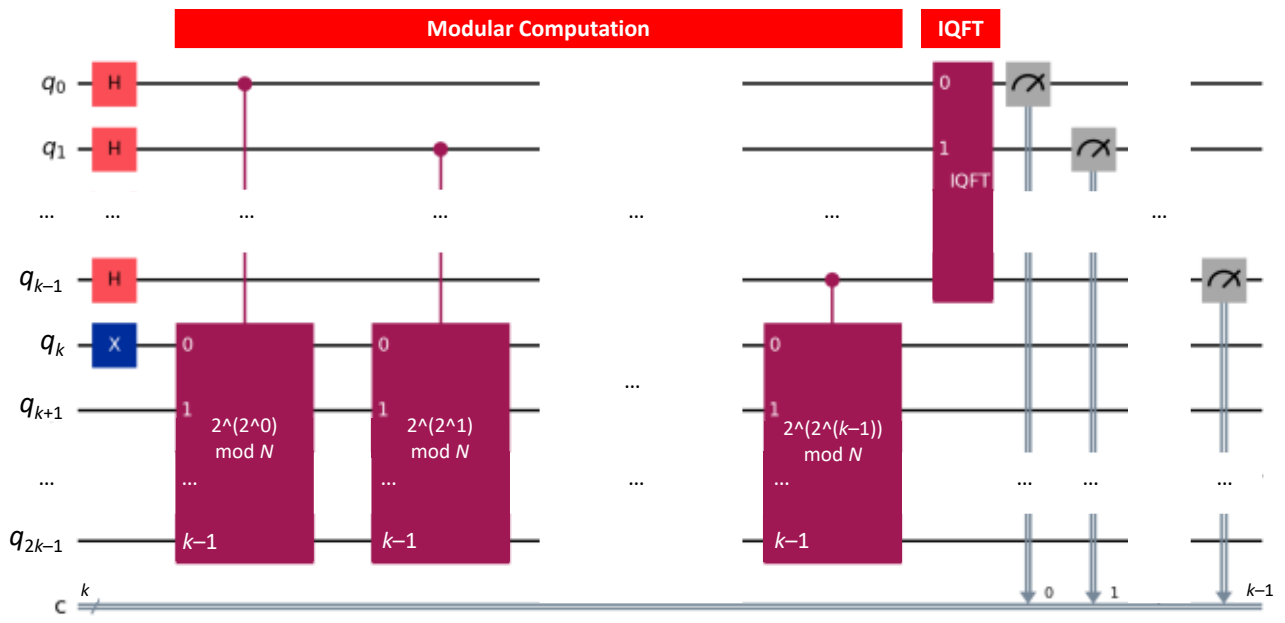

Fig. 1. Shor 演算法的基本流程

#### 1) 模數計算

在模數計算主要計算的值$f(i) = a^i(\mathrm{mod}\ N)$，其中$0 \le i \le 2^k - 1$，並且把結果紀錄到第 2 個註冊器。根據文獻[3]，模數計算結果存在子循環，符合$a^r(\mathrm{mod}\ N) \equiv 1(\mathrm{mod}\ N)$，所以表示也存在$a^j(\mathrm{mod}\ N) \equiv a^{j+r}(\mathrm{mod}\ N)$。

#### 2) 逆量子傅立葉變換計算

由前一個小節可知，模數計算後的值每隔週期 $r$ 會出現 1 次，所以每個模數計算後的值會呈現$\frac{1}{r}$的頻率。而在模數計算主要取得模數計算後的值，但 Shor 演算法的主要目標在於取得週期 $r$，所以需搭配逆量子傅立葉變換計算，可以在量子計算的特性下，用多項式計算時間取得每個計算結果的頻率和週期。

#### 3) 取得週期 r 和分解因數

通過逆量子傅立葉變換計算後進行量測，可以估計模數計算後的值的頻率和週期，從結果中可以取得週期 $r$。並且，通過公式(1)和公式(2)可以取得 $p$ 值和 $q$ 值。其中，$g$ 函數表示取最大公因數。需要注意的是，這個步驟在取得週期 $r$ 和分解因數的計算上是在經典電腦(classical computer)上執行，而不是量子電腦上執行。

$$p = g\left(a^{\frac{r}{2}} - 1, N\right). \quad (1)$$

$$q = g\left(a^{\frac{r}{2}} + 1, N\right). \quad (2)$$

*B.* 最先進的方法的模數計算量子電路

在實現 Shor 演算法上，IBM 提供了一個實作範例[5]，並且後續有許多文獻[6]-[10]都跟隨著 IBM 的範例來做嘗試。其中，在 IBM 提供的實作範例中，主要採用循環和 swap 的方式來迭代交換量子位元間的量子態作為模數計算的結果(即 **Algorithm 1** 的 **Line 02~ Line 04**)，如表 I 所示。迭代完成後，最後再轉換為控制邏輯閘(即 **Algorithm 1** 的 **Line 05**)，產生量子糾纏態。其中，當第 1 個註冊器的第 $b$ 個量子位元作為控制位元時，根據其量子態$|q_b\rangle$來執行 Algorithm 1 和產生第 2 個註冊器的量子位元量子態，則其迭代次數達$2^b k$。

TABLE I. 最先進的方法的模數計算量子電路[5]-[10]

**Algorithm 1.**最先進的方法的模數計算量子電路

```
01: Q = QuantumCircuit(k)
02: for _iteration in range(2^b):
03:   for i in range(k − 1):
04:     Q.swap(k − i − 2, k − i − 1)
05: Q = Q.control()
06: return Q
```

*C.* 常見案例—分解整數 15

目前已經有不少研究實作 Shor 演算法，並且多為採用分解整數 15 為例(即$N = 15$)來說明。其中，文獻[11]採用$a = 2$，、文獻[12]採用$a = 7$、文獻[13]採用$a = 13$，都表示可以順利分解整數 15 為 3 和 5 (即$p = 3$、$q = 5$)。有鑑於此，本節採用$a = 2$為例來說明分解整數 15，其 Shor 演算法的量子電路如圖 2 所示。

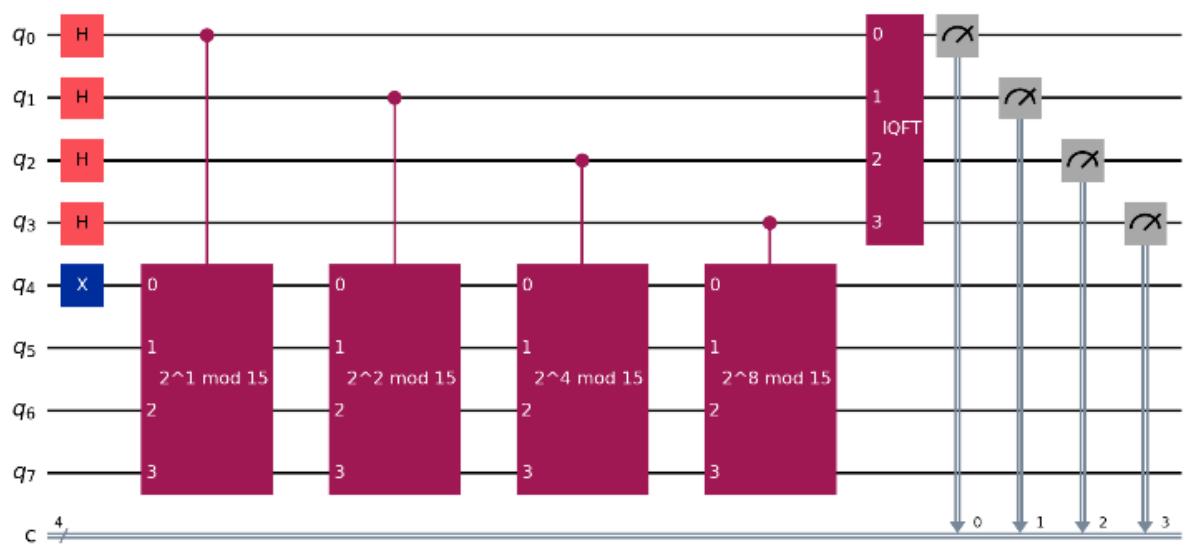

Fig. 2. 分解整數 15 的 Shor 演算法

*1)* 模數計算

根據第 II.A 節描述的模數計算方法，設定$k = 4$，首先將為每個量子位元操作 Hadamard Gate，讓其產生均勻疊加態，也就是可以同時展開$0 \sim 2^k - 1$的排列組合。在量子計算的特性下，可以操作 4 次 Algorithm 1 的量子電路來取得表 II 的模數計算結果，並且從可以觀察到模數計算結果只會有 4 種可能(即$|1\rangle$、$|2\rangle$、$|4\rangle$、$|8\rangle$)。這個結果也同時反應出週期$r = 4$。

*2)* 逆量子傅立葉變換計算

逆量子傅立葉變換計算主要用來計算每一個模數計算後的結果出現的頻率和週期。以表 II 結果，可以得到下面的結果：

- $|1\rangle(|0\rangle + |4\rangle + |8\rangle + |12\rangle)$
- $|2\rangle(|1\rangle + |5\rangle + |9\rangle + |13\rangle)$
- $|4\rangle(|2\rangle + |6\rangle + |10\rangle + |14\rangle)$
- $|8\rangle(|3\rangle + |7\rangle + |11\rangle + |15\rangle)$

由結果可知，每個模數計算後的結果出現的頻率為$\frac{4}{16} = \frac{1}{4}$，並且週期為 4。圖 3 為 Shor 演算法的量子電路執行後的量測結果。

TABLE II. 分解整數 15 的模數計算結果

| $i$ | $f(i)$ | 量子態$|q_{2k-1} \dots q_k\rangle|q_{k-1} \dots q_0\rangle = |f(i)\rangle|i\rangle$ |
|---|---|---|
| 0 | 1 | $|1\rangle|0\rangle$ |
| 1 | 2 | $|2\rangle|1\rangle$ |
| 2 | 4 | $|4\rangle|2\rangle$ |
| 3 | 8 | $|8\rangle|3\rangle$ |
| 4 | 1 | $|1\rangle|4\rangle$ |
| 5 | 2 | $|2\rangle|5\rangle$ |
| 6 | 4 | $|4\rangle|6\rangle$ |
| 7 | 8 | $|8\rangle|7\rangle$ |
| 8 | 1 | $|1\rangle|8\rangle$ |
| 9 | 2 | $|2\rangle|9\rangle$ |
| 10 | 4 | $|4\rangle|10\rangle$ |
| 11 | 8 | $|8\rangle|11\rangle$ |
| 12 | 1 | $|1\rangle|12\rangle$ |
| 13 | 2 | $|2\rangle|13\rangle$ |
| 14 | 4 | $|4\rangle|14\rangle$ |
| 15 | 8 | $|8\rangle|15\rangle$ |

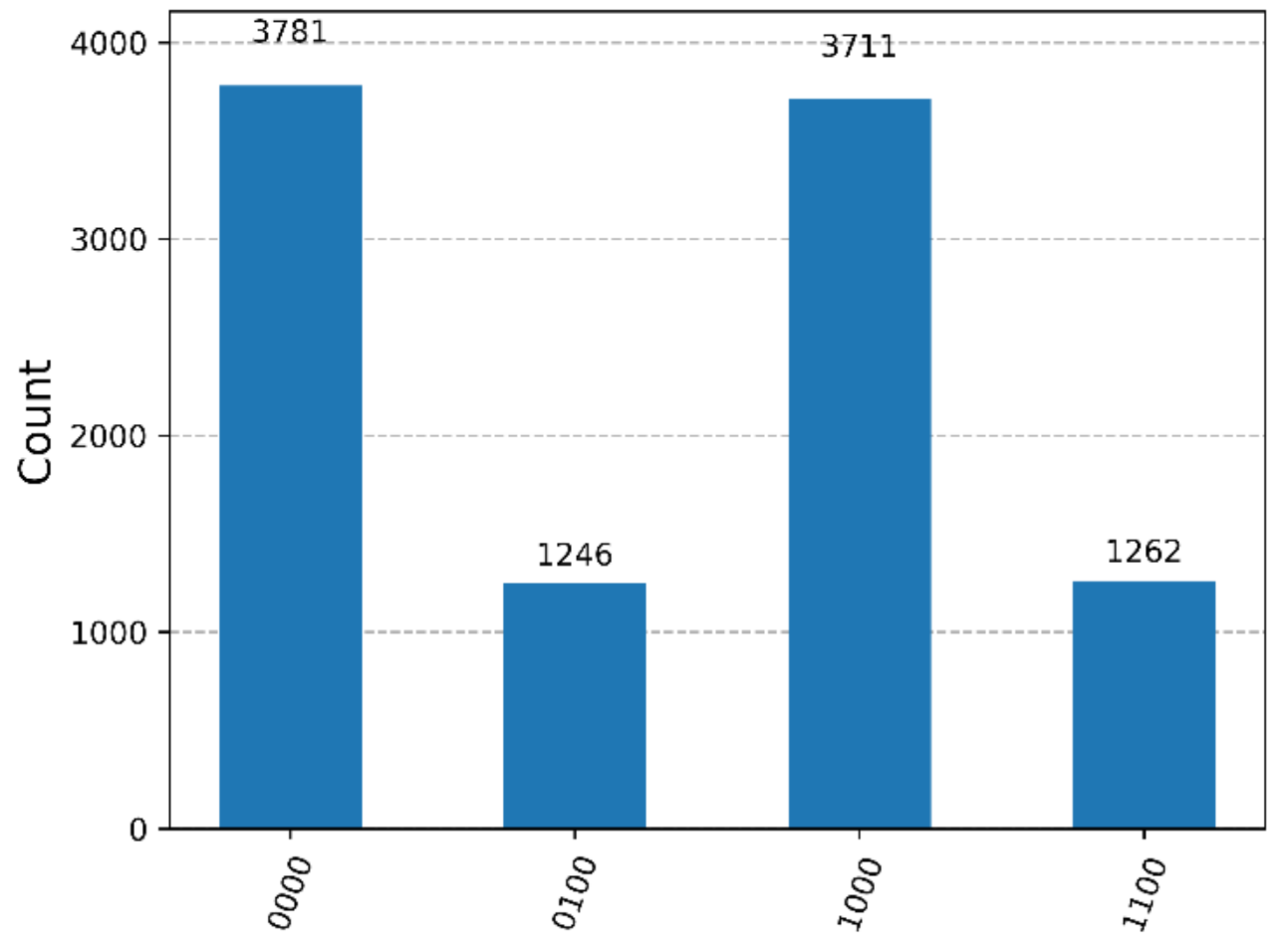

Fig. 3. 分解整數 15 的量子電路執行後的量測結果

*3)* *取得週期 r 和分解因數*

從逆量子傅立葉變換計算後進行量測，可以得到 4 個結果，其代表的意涵分別如下：

- $|0000\rangle \rightarrow 0 \rightarrow \frac{0}{16} \rightarrow 16$
- $|0100\rangle \rightarrow 4 \rightarrow \frac{4}{16} = \frac{1}{4} \rightarrow 4$
- $|1000\rangle \rightarrow 8 \rightarrow \frac{8}{16} = \frac{1}{2} \rightarrow 2$
- $|1100\rangle \rightarrow 12 \rightarrow \frac{12}{16} = \frac{3}{4} \rightarrow 4$

由結果可以觀察到 $r$ 有較高的可能性是 4，之後再採用公式(1)和公式(2)可以取得$p = g(3,15) = 3$和$q = g(5,15) = 5$。

## III. 本研究提出方法

本研究主要提出具有更高效率的模數計算量子電路，並且相較於最先進的方法的模數計算量子電路，可以用更少的量子位元和量子邏輯閘。然而，需要注意的是，本研究所提方法僅適用於$p = 3$，並且 $r$ 值為 2 的次方值。

### A. 本研究提出的模數計算量子電路

本研究提出的模數計算量子電路是採用經典電腦和量子電腦混合的作法，先在經典電腦計算$m = 2^b (\mathrm{mod}\ N)$ (即 **Algorithm 2** 的 **Line 01**)，之後再根據 $m$ 值來產生對應的模數值的量子電路(即 **Algorithm 2** 的 **Line 04~ Line 08**)，如表 III 所示。這個方法相較於最先進的方法的模數計算量子電路，可以少掉一層$2^b$循環，將可以大幅度減少計算量。

TABLE III. 本研究提出的模數計算量子電路

| **Algorithm 2.**本研究提出的模數計算量子電路 |
|---|
| **01:** $m = 2^b (\mathrm{mod}\ N)$ |
| **02:** $Q$ = QuantumCircuit($k$) |
| **03:** *isTwoPower* = false |
| **04:** for $i$ in range($k-1$): |
| **05:** if $m == 2^i$: |
| **06:** $Q$.swap(0, $i$) |
| **07:** *isTwoPower* = true |
| **08:** break |
| **09:** if *isTwoPower* == true: |
| **10:** $Q = Q$.control() |
| **11:** **return** $Q$ |
| **12:** else: |
| **13:** **return Algorithm 1** |

### B. 以分解整數 771 為例

在分解整數 771 時，如果採用最先進的方法的模數計算量子電路來計算，則$k = 16$。因此，由於最先進的方法的模數計算量子電路需要用到 32 個量子位元，但如果以 IBM Qiskit 實作時會需要執行 transpile 函數來建立量子電路，但 transpile 函數僅支援最多 31 個量子位元，所以將導致無法計算。

本研究提出的模數計算量子電路可以在$k = 12$情況下分解整數 771，結合本研究提出的模數計算量子電路的 Shor 演算法量子電路如圖 4 所示。

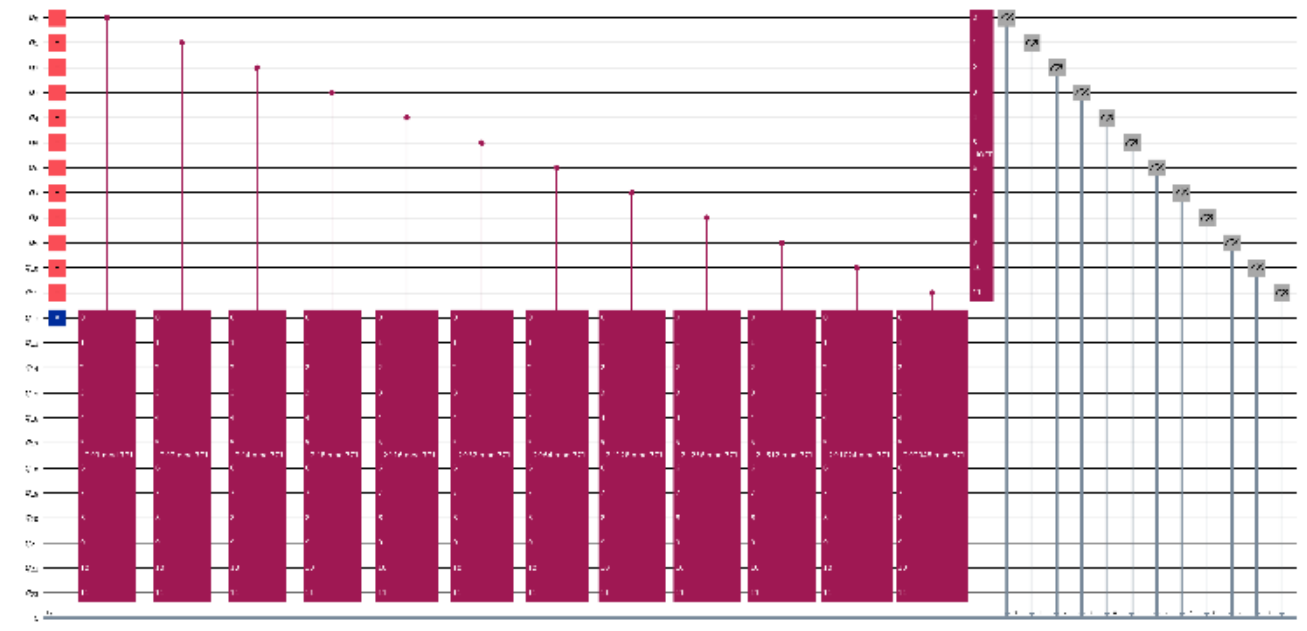

Fig. 4. 結合本研究提出的模數計算量子電路的 Shor 演算法

#### 1) 模數計算

根據第 III.A 節描述的模數計算方法，設定$k = 12$，首先將為每個量子位元操作 Hadamard Gate，讓其產生均勻疊加態，也就是可以同時展開$0 \sim 2^k - 1$的排列組合。在量子計算的特性下，可以操作 12 次 Algorithm 2 的量子電路來取得表 IV 的模數計算結果，並且從可以觀察到模數計算結果只會有 16 種可能(即$|1\rangle$、$|2\rangle$、$|4\rangle$、$|8\rangle$、$|16\rangle$、$|32\rangle$、$|64\rangle$、$|128\rangle$、$|256\rangle$、$|512\rangle$、$|253\rangle$、$|506\rangle$、$|241\rangle$、$|482\rangle$、$|193\rangle$、$|386\rangle$)。這個結果也同時反應出週期$r = 16$。

TABLE IV. 分解整數 771 的模數計算結果

| $i$ | $f(i)$ | 量子態$\lvert q_{2k-1} \dots q_k\rangle \lvert q_{k-1} \dots q_0\rangle = \lvert f(i)\rangle \lvert i\rangle$ |
|---|---|---|
| 0 | 1 | $\lvert 1\rangle \lvert 0\rangle$ |
| 1 | 2 | $\lvert 2\rangle \lvert 1\rangle$ |
| 2 | 4 | $\lvert 4\rangle \lvert 2\rangle$ |
| 3 | 8 | $\lvert 8\rangle \lvert 3\rangle$ |
| | | … |
| 4095 | 8 | $\lvert 386\rangle \lvert 4095\rangle$ |

#### 2) 逆量子傅立葉變換計算

逆量子傅立葉變換計算主要用來計算每一個模數計算後的結果出現的頻率和週期。以表 IV 結果，可以得到下面的結果：

- $|1\rangle(|0\rangle + |16\rangle + |32\rangle + \cdots + |4080\rangle)$
- $|2\rangle(|1\rangle + |17\rangle + |33\rangle + \cdots + |4081\rangle)$
- $|4\rangle(|2\rangle + |18\rangle + |34\rangle + \cdots + |4082\rangle)$
- $|8\rangle(|3\rangle + |19\rangle + |35\rangle + \cdots + |4083\rangle)$
- …
- $|386\rangle(|15\rangle + |31\rangle + |47\rangle + \cdots + |4095\rangle)$

由結果可知，每個模數計算後的結果出現的頻率為$\frac{256}{4096} = \frac{1}{16}$，並且週期為 16。圖 5 為 Shor 演算法的量子電路執行後的量測結果。

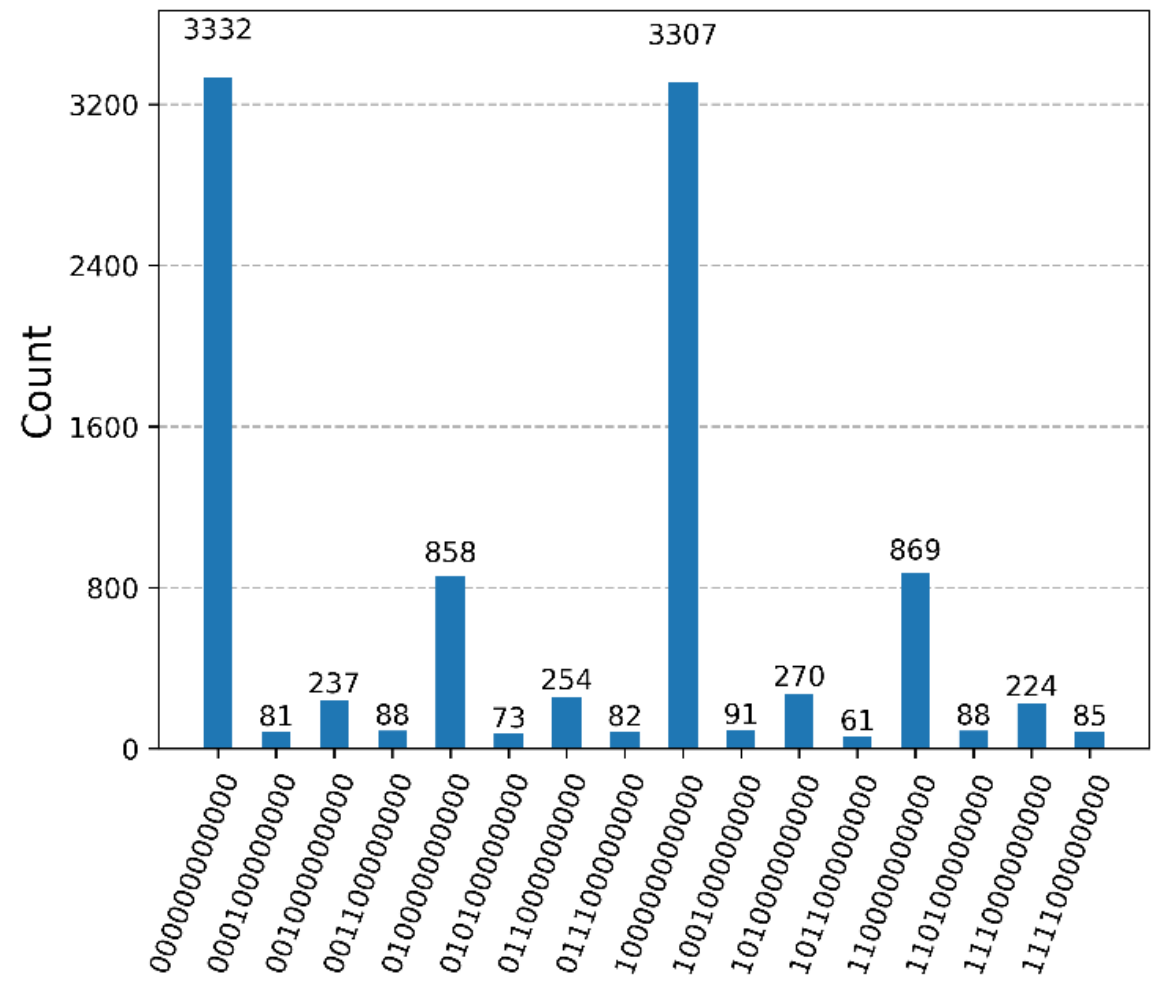

Fig. 5. 分解整數 771 的量子電路執行後的量測結果

#### 3) 取得週期 r 和分解因數

從逆量子傅立葉變換計算後進行量測，可以得到 16 個結果，其代表的意涵分別如下：

- $|000000000000\rangle \to 0 = \frac{0}{4096} \to 4096$
- $|000100000000\rangle \to 256 \to \frac{256}{4096} = \frac{1}{16} \to 16$
- $|001000000000\rangle \to 512 \to \frac{512}{4096} = \frac{1}{8} \to 8$
- $|001100000000\rangle \to 768 \to \frac{768}{4096} = \frac{3}{16} \to 16$

- …
- $|111100000000\rangle \rightarrow 3840 \rightarrow \frac{3840}{4096} = \frac{15}{16} \rightarrow 16$

由結果可以觀察到$r$有較高的可能性是 16，之後再採用公式(1)和公式(2)可以取得$p = g(3,771) = 3$和$q = g(257,771) = 257$。

## IV. 實驗結果與討論

在 IV.A 節將介紹本研究的實驗環境，並且在 IV.B 節討論實驗結果，比較本研究提出的方法和最先進的方法的結果。

### *A. 實驗環境*

本研究實驗時所採用的硬體和軟體規格描述如下：CPU Intel(R) Core(TM) i7-10510U、RAM 16 GB、Python 3.11.9、NumPy 1.23.5、以及 Qiskit 1.1.1。其中，在量子計算的部分，主要採用 Qiskit 的 AER Simulator 進行模擬。為驗證本研究提出的方法，總共採用 12 個案例來進行評估與比較，每個案例在執行 Shor 演算法時皆進行 1 萬次 shot 量測，案例的詳細參數值描述於附錄 A。

### *B. 結果比較與討論*

在本研究中主要採用附錄 A 描述的 12 個案例來比較本研究提出的方法和最先進的方法[5]-[10]，並且分別從使用的量子位元數、量子電路產製時間長度、以及 Shor 演算法執行時間長度來進行比較。

表 V 為 Shor 演算法使用的量子位元數及其產製時間長度，其時間單位為秒。其中，由於 IBM Qiskit 1.1.1 的 transpile 函數僅支援最多 31 個量子位元，所以如果量子電路超過該限制將無法被執行。因此，採用最先進的方法僅能實作 Case 1 和 Case 2，而其他 10 個案例將標記為“ Not applicable”。從實驗結果可以發現，本研究提出的模數計算量子電路可以用更少的量子位元來分解出4096-bit 整數。

TABLE V. SHOR 演算法使用的量子位元數及其產製時間長度

| Case | The Number of Qubits | | Quantum Circuit Generation Time (s) | |
|---|---|---|---|---|
| | The Proposed Method | The SOTA Methods [5]-[10] | The Proposed Method | The SOTA Methods [5]-[10] |
| **Case 1** | 8 | 8 | 0.011 | 0.029 |
| **Case 2** | 16 | 16 | 0.020 | 0.576 |
| **Case 3** | 24 | 32 | 0.048 | Not applicable |
| **Case 4** | 24 | 64 | 0.120 | Not applicable |
| **Case 5** | 24 | 128 | 0.250 | Not applicable |
| **Case 6** | 24 | 256 | 0.508 | Not applicable |
| **Case 7** | 24 | 512 | 0.873 | Not applicable |
| **Case 8** | 24 | 1024 | 2.048 | Not applicable |
| **Case 9** | 24 | 2048 | 3.970 | Not applicable |
| **Case 10** | 24 | 4096 | 14.696 | Not applicable |
| **Case 11** | 24 | 8192 | 33.467 | Not applicable |
| **Case 12** | 26 | 16384 | 68.334 | Not applicable |

除此之外，在 Case 1 採用相同量子位元數的情況，最先進的方法的產製量子電路時間長度是本研究提出方法的產製量子電路時間長度 2.7 倍；相似的，Case 2 採用相同量子位元數的情況，最先進的方法的產製量子電路時間長度是本研究提出方法的產製量子電路時間長度 28.7 倍。有鑑於此，即使 transpile 函數可支援最先進的方法的量子位元數，但最先進的方法的量子電路產製時間也將呈指數增加。

表 VI 為 Shor 演算法使用的量子位元數及其執行時間長度，其單位時間為秒。由於本研究執行 Shor 演算法將進行 1 萬次 shot 量測，所以將需要花費較多的時間在執行和量測。隨著分解的整數值越大，則需要的執行時間長度則會越大。以 Case 11 來觀察，可以發現採用本研究提出的方法可以在 1 小時內分解 4096-bit 整數。

TABLE VI. SHOR 演算法使用的量子位元數及其執行時間長度

| Case | The Number of Qubits | | Quantum Circuit Execution Time (s) | |
|---|---|---|---|---|
| | The Proposed Method | The SOTA Methods [5]-[10] | The Proposed Method | The SOTA Methods [5]-[10] |
| **Case 1** | 8 | 8 | 0.050 | 0.112 |
| **Case 2** | 16 | 16 | 0.077 | 2.132 |
| **Case 3** | 24 | 32 | 0.152 | Not applicable |
| **Case 4** | 24 | 64 | 5.469 | Not applicable |
| **Case 5** | 24 | 128 | 14.628 | Not applicable |
| **Case 6** | 24 | 256 | 34.696 | Not applicable |
| **Case 7** | 24 | 512 | 78.052 | Not applicable |
| **Case 8** | 24 | 1024 | 171.513 | Not applicable |
| **Case 9** | 24 | 2048 | 360.163 | Not applicable |
| **Case 10** | 24 | 4096 | 733.837 | Not applicable |
| **Case 11** | 24 | 8192 | 2229.684 | Not applicable |
| **Case 12** | 26 | 16384 | 11308.347 | Not applicable |

## V. 結論與未來研究

本研究提出高效率的模數計算量子電路，與最先進的方法[5]-[10]比較，證明可以提升二十倍的效率以上。除此之外，本研究提出的方法可以用更少的量子位元數來分解長度更長的整數，包含分解 4096-Bit 整數。在實驗環境中，本研究採用 IBM Qiskit 實作量子電路和量子演算法，通過實作來驗證本研究提出方法可以在 1 小時內分解 4096-Bit 整數。然而，需要注意的是，本研究所提方法僅適用於$p = 3$，並且$r$值為 2 的次方值之案例。

由於本研究方法僅限於特定條件限制下分解 4096-Bit 整數，距離通用型模數計算量子電路仍有段距離。在未來研究中可以嘗試設計高效率的通用型模數計算量子電路，讓量子計算更能解決各種應用問題。

## 致謝

本研究承蒙 IBM 協助免費提供 IBM Quantum Platform 和 IBM Qiskit，讓本研究得以實作 Shor 演算法進行實驗和分析，特此感謝 IBM 提供 Open Plan。

## 參考文獻

## APPENDIX A

為驗證本研究提出的方法，設計 12 個案例的 $p$ 值和 $q$ 值，其中$N = pq$。Case 1 為常見的案例，分解 15，得到其質因數分別有 3 和 5。另外 11 個 Case 為長度更長的整數，並且被本研究方法被分解出來。

### 1) *Case 1 for Factoring a 4-Bit Integer*

$p$: 3

$q$: 5

$N$: 15

### 2) *Case 2 for Factoring a 8-Bit Integer*

$p$: 3

$q$: 17

$N$: 51

### 3) *Case 3 for Factoring a 16-Bit Integer*

$p$: 3

$q$: 257

$N$: 771

### 4) *Case 4 for Factoring a 32-Bit Integer*

$p$: 3

$q$: 65537

$N$: 196611

### 5) *Case 5 for Factoring a 64-Bit Integer*

$p$: 3

$q$: 4294967297

$N$: 12884901891

### 6) *Case 6 for Factoring a 128-Bit Integer*

$p$: 3

$q$: 18446744073709551617

$N$: 55340232221128654851

### 7) *Case 7 for Factoring a 256-Bit Integer*

$p$: 3

$q$: 340282366920938463463374607431768211457

$N$: 1020847100762815390390123822295304634371

### 8) *Case 8 for Factoring a 512-Bit Integer*

$p$: 3

$q$: 115792089237316195423570985008687907853269984665640564039457584007913129639937

$N$: 347376267711948586270712955026063723559809953996921692118372752023739388919811

### 9) *Case 9 for Factoring a 1024-Bit Integer*

$p$: 3

$q$: 13407807929942597099574024998205846127479365820592393377723561443721764030073546976801874298166903427690031858186486050853753882811946569946433649006084097

$N$: 40223423789827791298722074994617538382438097461777180133170684331165292090220640930405622894500710283070095574559458152561261648435839709839300947018252291

### 10) *Case 10 for Factoring a 2048-Bit Integer*

$p$: 3

$q$: 179769313486231590772930519078902473361797697894230657273430081157732675805500963132708477322407536021120113879871393357658789768814416622492847430639474124377767893424865485276302219601246094119453082952085005768838150682342462881473913110540827237163350510

684586298239947245938479716304835356329624224137217

$N$: 539307940458694772318791557236707420085393093682691971820290243473198027416502889398125431967222608063360341639614180072976369306443249867478542291918422373133303680274596455828906658803738282358359248856255017306514452047027388644421739331622481711490051532053758894719841737815439148914506068988872672411651

### 11) *Case 11 for Factoring a 4096-Bit Integer*

$p$: 3

$q$: 3231700607131100730071487668866995196044410266971548403213034542752465513886789089319720141152291346368871796092189801949411955915049092109508815238644828312063087736730099609175019775038965210679605763838406756827679221864261975616183809433847617047058164585203630504288757589154106580860755239912393038552191433338966834242068497478656456949485617603532632205807780565933102619270846031415025859286417711672594360371846185735759835115230164590440369761323328723122712568471082020972515710172693132346967854258065669793504599726835299863821552516638943733554360213543322960464531847860495214819355585361105959623065 7

$N$: 969510182139330219021446300660098558813323080091464520963910362825739654166036726795916042345687403910661538827656940584823586774514727632852644571593448493618926321019029882752505932511689563203881729151522027048303766559278592684855142830154285114117449375561089151286627276746231974258226571973717911565657430001690050272620549243596937084845685281059789661742334169779930785781253809424507757785925313501778308111553855720727950534569049377132110928396998616936813770541324606291754713051807939704090356277419700938051379918050589959146465754991683120066308064062996888139359554358148564445806675608331787886919 71

### 12) *Case 12 for Factoring a 8192-Bit Integer*

$p$: 3

$q$: 1044388881413152506691752710716624382579964249047383780384233483283953907971557456848826811934997558340890106714439262837987573438185793607263236087851365277945956976543709998340361590134383718314428070011855946226376318839397712745672334684344586617496807908705803704071284048740118609114467977783598029006686938976881787785946905630190260940599579453432823469303026696443059025015972399867714215541693835559885291486318237914434496734087811872639496475100189041349008417061675093668333850551032972088269550769983616369411933015213796825837188091833656751221318492846368125550225998300412344784862595674492194617023806505913245610825731835380087608622102834270197698202313169017678006675195485079921636419370285375124784014907159135459982790513399611551794271106831134090584272884279791554849782954323534517065223269061394905987693002122963395687782878948440616007412945674919823050571642377154816321380631045902916136926708342856440730447899971901781465763473223850267253059899795996090799469201774624817718449867455659250178329070473119433165550807568221846571746373296884912819520317457002440926616910874148385078411929804522981857338977648103126085903001302413467189726673216491511131602920781738033436090243804708340403154190337

$N$: 3133166644239457520075258132149873147739892747142151341152700449851861723914672370546480435804992675022670320143317788513962720314557380821789708263554095833837870929631129995021084770403151154943284210035567838679128956518193138237017004053033759852490423726117411112213852146220355827343403933350794087020060816930645363357840716890570782821798738360298470407909080089329177075047917199603142646625081506679655874458954713743303490202263435617918489425300567124047025251185025281005001551653098916264808652309950849108235799045641390477511564275500970253663955478539104376650677994901237034354587787023476583851071419517739736832477195506140262825866308502810593094606939507053034020025586455239764909258110856125374352044721477406379948371540198834655382813320493402271752818652839374664549348862970603551195669807184184717963079006368890187063348636845321848022238837024759469151714927131464448964141893137708748410780125028569322191343699915705344397290419671550801759179699387988272398407605323874453155349602366977750534987211419358299496652422704665539715239119890654738458560952371007322779850732622445155235235789413568945572016932944309378257709003907240401569180019649474533394808762345214100308270731414125021209462571011